\documentclass[sigconf,nonacm]{acmart}

\usepackage{hyperref}
\begin{document}

\title{Triplet: Triangle Patchlet for Mesh-Based Inverse Rendering and Scene Parameters Approximation} 
\author{Jiajie Yang}
\authornote{ This is a \textbf{draft} which only present the main idea of the paper. Up-to-date experiment code of this paper is available at \textbf{\href{https://github.com/RANDO11199/Triplet}{GitHub}}. The plan for update is also published on the page. If you have any question, please contact me via email: jiajie.y@wustl.edu.
}
\authornote{Permission to make digital or hard copies of all or part of this work for personal or
 classroom use is granted without fee provided that copies are not made or distributed
 for profit or commercial advantage and that copies bear this notice and the full citation
 on the first page. Copyrights for third-party components of this work must be honored.
 For all other uses, contact the owner/author(s).
 ©2024 Copyright held by the owner/author(s).}
\email{jiajie.y@wustl.edu}
\renewcommand{\shortauthors}{Yang}

\begin{abstract}
 Recent advancements in Radiance Fields have significantly improved novel-view synthesis. However, in many real-world applications, the more advanced challenge lies in inverse rendering, which seeks to derive the physical properties of a scene, including light, geometry, textures, and materials. Meshes, as a traditional representation adopted by many simulation pipeline, however, still show limited influence in radiance field for inverse rendering. This paper introduces a novel framework called Triangle Patchlet (abbr. Triplet), a mesh-based representation, to comprehensively approximate these scene parameters. We begin by assembling Triplets with either randomly generated points or sparse points obtained from camera calibration where all faces are treated as an independent element. Next, we simulate the physical interaction of light and optimize the scene parameters using traditional graphics rendering techniques like rasterization and ray tracing, accompanying with density control and propagation. An iterative mesh extracting process is also suggested, where we continue to optimize on geometry and materials with graph-based operation. We also introduce several regulation terms to enable better generalization of materials property. Our framework could precisely estimate the light, materials and geometry with mesh without prior of light, materials and geometry in a unified framework. Experiments demonstrate that our approach can achieve state-of-the-art visual quality while reconstructing high-quality geometry and accurate material properties.
\end{abstract}


\begin{teaserfigure}
  \includegraphics[width=\textwidth]{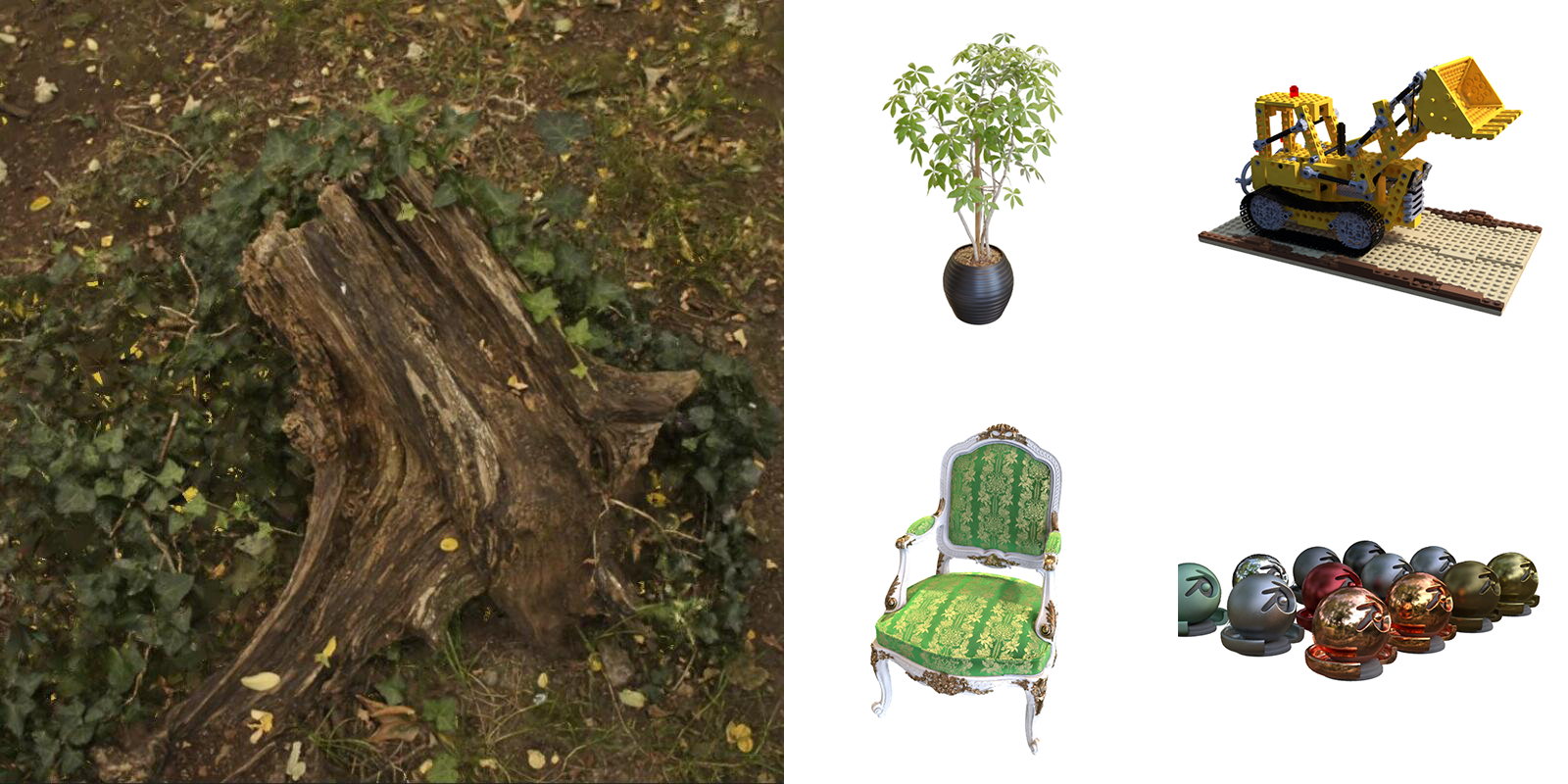}
  \caption{Triplet utilizes a traditional mesh-based rendering pipeline to extract physically-based materials, geometry, and lighting from scenes with complex topology, all within a unified framework, while delivering excellent visual quality.}
  \label{fig:teaser}
\end{teaserfigure}

\maketitle

\section{Introduction}

The rise of AI-generated content (AIGC) has provided a powerful and convenient tool for 3D production. Many researchers focus on novel-view synthesis \cite{nerf}\cite{gs3d} for scene reconstruction. However, in applications such as gaming, movies, and autonomous driving, more precise scene parameters are required. These extend beyond view synthesis and fall into the domain of inverse rendering, which involves deriving physical properties—such as geometry, materials, and lighting—from 3D or 2D images. While many representations have achieved success in novel-view synthesis, meshes are widely adopted as a conventional representation for simulation across various fields. Meshes are used extensively in areas like solid modeling, video games, virtual reality (VR), augmented reality (AR), and virtual avatars. However, many of these representations are not fully compatible with traditional computer graphics pipelines, and achieving subtle congruence with physical reality requires additional effort in real-world applications for those scenario where precise scene parameters is required. In such cases, separating physically plausible elements like lighting, geometry, and materials is preferred. Furthermore, modern computer graphics rendering pipelines are predominantly mesh-based, offering a well-researched foundation for achieving physically-based rendering (PBR), which is crucial for recovering physical properties from scenes.

This paper proposes a novel approach to inverse rendering. First, we use small triangle faces to represent the local properties of a surface and optimize collections of discrete meshes, ultimately refining the parameters of the entire scene. We perform a dense covering with broken and discrete meshes that gradually conform to the potential surface using gradient flow. However, a overlapped covering introduces optimization challenges, which will be discussed in detail later. Additionally, we treat the optimization process as a simulation, where scene parameters are iteratively optimized using physically plausible methods. The accuracy of the final result largely depends on the physical processes followed during the simulation.

This paper aims to overcome several obstacles when using mesh as representation for inverse rendering and try to enable direct optimization of meshes in terms of both geometry and materials, using simulation techniques from traditional computer graphics, such as rasterization and ray tracing. This approach allows for optimization in reasonable time, either offline or online, for scenes or assets captured from multi-view photos.

To achieve our goals, the framework must overcome four major challenges. The first challenge is optimizing the mesh for objects and scenes with complex geometries. A common optimization approach involves deforming a mesh initialized from a sphere, treating it as a closed graph or geometry must be given. For mesh-based method required only multi-view RGB image as input, the reconstructed object is still simple and potential improve on quality is existed. These method struggles with objects and scenes that have complex topological characteristic as well as the unknown light condition. Recognizing the dual nature of triangle meshes—where they represent both the overall topological relation and the local optical properties —we break the mesh into small triangle patch to address this issue by approximating the local property of surface with small patch at the very beginning and optimizing the mesh with refining on iteratively extracted mesh.

The second challenge is how to propagate gradients throughout the entire scene. Neural Radiance Fields (NeRF) naturally handle this, and some works combine them with other methods \cite{vosh}. However, traditional z-buffer algorithms hinder gradient propagation because only one triangle per pixel is typically selected to compute the final color value \cite{DRsurvey}. As a result, basic elements may not cover the correct pixels, and gradients cannot propagate to the optimal elements. This issue affects all methods that approximate scene geometry with small, compact elements. Simply no gradient will be propagate to if no elements cover the potential surface from the camera view. To address this problem, we treat the mesh as a set of "colored glass" patches, where light can be absorbed, reflected, or transmitted, and multiple triangles at each pixel are blended. We will discuss with more details in Section.\ref{sec3}. Also, we initialize the process with sparse or random points, combined with density control and gradients, to densely cover the surface with triangle meshes.  We also gradually refine the meshes through an iterative surface extraction process, accompanying with dynamic density control of local topology connection for extracted meshes.

The third major challenge is handling light from the environment and between object. Lighting plays a crucial role in determining the scene's textures, materials, and geometry especially when reconstructing object with reflection effect. Optimization could go into local optimum where the textures/materials are overfitting and compensate the light transport deviating from reality. For objects with simple lighting conditions or bounded scene, we use point lights, while directional lights are employed for outdoor scenes. Multiple light sources require multiple evaluations, but this can result in overly comprehensive approximations. Therefore, for fast query of comprehensive light including direct and indirect light at specific vertex, we introduce vertex-based spherical harmonics to estimate the average radiance and direction received from the hemisphere, which help separating lighting and materials without the contamination of materials/textures potentially incurred by the ignorance of indirect light, especially for rasterization.

The fourth challenge lies in the rendering pipeline. To balance fidelity and efficiency, we adopt both Blinn-Phong and Cook-Torrance reflection models. Blinn-Phong, an empirical model with fast rendering speed, is physically plausible and is used in our rasterization pipeline. However, Blinn-Phong does not strictly adhere to the law of conservation of energy. For a physically accurate simulation with reasonable computational cost, we use the Cook-Torrance bidirectional reflectance distribution function (BRDF) as our reflection model. To balance the convenience of shader editing, speed and memory consumption especially for Physically-based Rendering, we implemented the shader framework with a combination of Cuda and PyCuda.

To summarize, this paper: 
\begin{itemize} 
\item Introduces Triplet, a mesh-based framework for reconstructing scenes and objects with complex topologies using sparse or random initialization, and only RGB images as input, for the first time to the best of our knowledge. 
\item Provides a novel method for estimating materials, textures, lighting, and geometry of a scene in one unified process. 
\item Suggests a differentiable rendering pipeline for extracting the physical parameters of a scene without any prior knowledge of environmental lighting. 
\item Proposes an optimization method for Triplet, incorporating adaptive density control, iterative mesh extraction, and regularization terms. This allows Triplet to capture and propagate the physical properties and detailed geometry of unknown surfaces, while resulting in high-quality visual effects. 
\item Offers a CUDA/PyCuda implementation for fast computation and flexible user customization. 
\end{itemize}

\section{Rendering Pipeline}
The most fundamental and important factor for extraction a correct geometry and materials from RGB images is the rendering pipeline. Rendering pipeline could be separate into two part. One is how we simulate the interaction between light, geometry and materials. We simulate with Blinn-Phong reflection model and Cook-Torrance reflection in this paper. The other one is how to present these interactions on screen. We use rasterization and ray tracing techniques in this paper.
\subsection{Reflection Model}
The rendering equation\cite{kajiya1986rendering} is given by:
\begin{equation}
L_o(\mathbf{x}, \mathbf{w}) = L_e(\mathbf{x}, \mathbf{w}) + \int_{H} f_r(\mathbf{x}, \mathbf{w}, \mathbf{w}') L_i(\mathbf{x}, \mathbf{w}') (\mathbf{n} \cdot \mathbf{w}') d\mathbf{w}'
\end{equation}. The rendering could consider into two part, which is separately the light transport and the interaction between surface and light.

\textbf{Blinn-Phong Reflection Model}. Blinn-Phong reflection model\cite{blinn1977models}\cite{phong1975illumination} is utilized extensively for simulating the way light interacts with surfaces to produce realistic shading effects. Blinn-Phong Model decomposes the light reflected from a surface into three main components: ambient, diffuse and specular reflection:
\begin{align}
    f_r(\mathbf{x}, \mathbf{w}, \mathbf{w}') =  k_{d} \cdot max(0,N \cdot L) + k_{s} \cdot max(0,N \cdot H)^s ,
\end{align}where $k_d$ is diffuse coefficients of the surface $I_L$ is the intensity of the light source, $N$ is the noramlized surface normal and L is the normalized vector pointing towards the light source, $k_s$ is the specular coefficients of the surface and H is the normalized half-vector between light direction and view direction, $s$ is the shininess coefficient. Blinn-Phong model efficiently simulates a wide range of surface appearances, from matte to highly reflective. Its balance of computational efficiency and visual fidelity has made it a widely adopted technique in various fields, including video game development, simulations, and visual effects. However, Blinn-Phong model cannot simulate many surface effect, such as glossy reflection or anisotropic appearance.

\section{Rendering Pipeline}
The most fundamental factor for extracting accurate geometry and materials from RGB images is the rendering pipeline. The rendering pipeline can be divided into two parts. The first part involves simulating the interaction between light, geometry, and materials. In this paper, we use both the Blinn-Phong and Cook-Torrance reflection models. The second part involves presenting these interactions on the screen, where we employ rasterization and ray tracing techniques.

\subsection{Reflection Models}
The rendering equation \cite{kajiya1986rendering} is given by: \begin{equation} L_o(\mathbf{x}, \mathbf{w}) = L_e(\mathbf{x}, \mathbf{w}) + \int_{H} f_r(\mathbf{x}, \mathbf{w}, \mathbf{w}') L_i(\mathbf{x}, \mathbf{w}') (\mathbf{n} \cdot \mathbf{w}') d\mathbf{w}' \end{equation} The rendering process can be considered in two parts: light transport and the interaction between surfaces and light.

\textbf{Blinn-Phong Reflection Model}
The Blinn-Phong reflection model \cite{blinn1977models} \cite{phong1975illumination} is widely used to simulate how light interacts with surfaces, producing realistic shading effects. The model decomposes the light reflected from a surface into three components: ambient, diffuse, and specular reflection: \begin{align} f_r(\mathbf{x}, \mathbf{w}, \mathbf{w}') = k_{d} \cdot \max(0, N \cdot L) + k_{s} \cdot \max(0, N \cdot H)^s \end{align} where $k_d$ is the diffuse coefficient of the surface, $k_s$ is the specular coefficient, $N$ is the normalized surface normal, $L$ is the normalized vector pointing towards the light source, and $H$ is the normalized half-vector between the light direction and the view direction. The shininess coefficient $s$ controls the sharpness of the specular reflection. The Blinn-Phong model effectively simulates a range of surface appearances, from matte to highly reflective. Due to its balance of computational efficiency and visual fidelity, it is commonly used in video game development, simulations, and visual effects. However, the Blinn-Phong model cannot accurately simulate effects like glossy reflections or anisotropic surfaces.

\textbf{Cook-Torrance Reflection Model}
The Cook-Torrance reflection model offers a more physically accurate representation of light interaction with surfaces by incorporating the microfacet structure that defines the appearance of materials. It takes into account Fresnel effects, geometric attenuation, and microfacet distribution, making it suitable for rendering complex materials like metals and plastics. The Bidirectional Reflectance Distribution Function (BRDF) in the Cook-Torrance model is expressed as: \begin{equation} f_r(\mathbf{x}, \mathbf{w}, \mathbf{w}') = \frac{FGD}{4(\mathbf{N} \cdot \mathbf{L})(\mathbf{N} \cdot \mathbf{V})} \end{equation} where $F$ represents the Fresnel effect, $G$ is the geometric attenuation term, and $D$ is the microfacet distribution function.

We use the Fresnel-Schlick approximation for $F$: \begin{equation} F = F_0 + (1 - F_0) \cdot (1 - \cos(\theta))^5 \end{equation} For the geometric term, we use the Schlick-GGX approximation: \begin{equation} G_{\text{Schlick-GGX}}(\mathbf{N}, \mathbf{V}, k) = \frac{\mathbf{N} \cdot \mathbf{V}}{(\mathbf{N} \cdot \mathbf{V})(1 - \frac{(r + 1)^2}{8}) + \frac{(r + 1)^2}{8}} \end{equation} The Smith approximation for both view and light directions is: \begin{equation} G_{\text{Smith}}(\mathbf{N}, \mathbf{V}, \mathbf{L}, k) = G_{\text{Schlick-GGX}}(\mathbf{N}, \mathbf{V}, r) \cdot G_{\text{Schlick-GGX}}(\mathbf{N}, \mathbf{L}, r) \end{equation} The microfacet distribution $D$ is modeled using the Trowbridge-Reitz/GGX distribution \cite{trowbridge1975average} \cite{GGXwalter2007microfacet}: \begin{equation} D = \frac{\alpha^2}{\pi \left((\mathbf{N} \cdot \mathbf{H})^2 (\alpha^2 - 1) + 1\right)^2} \end{equation} To ensure energy conservation, we integrate the Fresnel term into the specular component $k_s$, while $k_d$ represents the refracted energy. The final BRDF is: \begin{equation} f(\mathbf{x}, \mathbf{w}, \mathbf{w}') = f_d + f_r = (1 - F) \frac{k_d}{\pi} + F \frac{G D}{4(\mathbf{N} \cdot \mathbf{L})(\mathbf{N} \cdot \mathbf{V})} \end{equation}

The Cook-Torrance model is widely used in computer graphics for real-time physically based rendering (PBR). While it offers high realism, it is computationally more expensive compared to simpler models like Blinn-Phong.

\subsection{Light Transport}
One of the most challenging and important steps in accurately reconstructing scene parameters is modeling the process of light transport. In this work, we adopt point lights, directional lights, and spherical harmonics lighting, each with learnable parameters, to simulate various lighting conditions. We evaluate the effectiveness of both rasterization and ray tracing for light transport.

\textbf{Light Sources}
In this paper, we use point lights, directional lights, and spherical harmonics for lighting. For physically-based rendering, point lights follow the inverse-square law, which provides a good approximation for reconstructing objects or indoor scenes with bounded geometry. For outdoor scenes, directional lights or point lights without the inverse-square law are used as reasonable approximations in most cases. However, these approaches offer coarse approximations of environmental lighting, which can limit realism.

To address this, we introduce learnable Spherical Harmonics (SH) lighting to model complex environment light using Image-Based Lighting (IBL) techniques or as an approximation of direct irradiance. For direct lighting using spherical harmonics, we query the overall light direction and the radiance received from the hemisphere at each vertex location, which enabling reflection between object even without ray tracing while separating well between lights and materials.

\textbf{Rendering Methods}
Rasterization and ray tracing are two fundamental techniques in rendering images for computer graphics, each with distinct advantages and limitations.

Rasterization is faster and well-suited for real-time applications. However, it processes each triangle independently, limiting its ability to simulate complex light interactions like global illumination. Reflections and refractions are often approximated using techniques such as cube maps or screen space reflections (SSR), which are less accurate. Shadows are typically approximated using shadow maps, which can also lack precision.

Ray Tracing, on the other hand, is slower due to the need for tracing multiple rays per pixel, testing for intersections, and simulating light interactions such as reflections, refractions, and shadows. However, it excels at handling multiple light bounces, producing highly accurate and realistic reflections and refractions. Ray tracing also generates soft shadows and penumbras naturally, with proper light falloff and occlusion.

In summary, rasterization offers superior performance while sacrificing accuracy, making it ideal for real-time applications. In contrast, ray tracing prioritizes visual fidelity but at a higher computational cost, which is better suited for offline rendering. This paper explores and tests both methods to balance efficiency and quality depending on the specific requirements. We also implement a mixed solution which combine both rasterization and ray tracing.

\begin{figure*}[!h]
	\includegraphics[width=\linewidth]{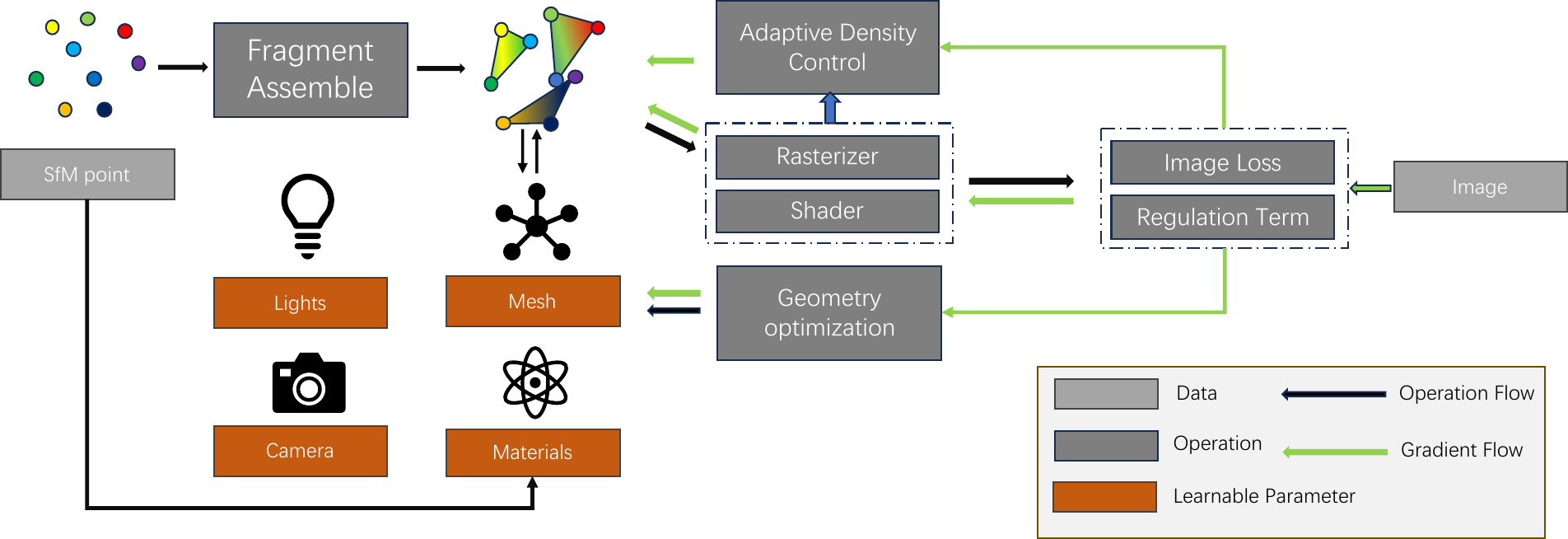}
	\caption{
		Optimization starts with the sparse SfM point cloud and creates a set of 3D Gaussians. We then optimize and adaptively control the density of this set of Gaussians. During optimization we use our fast tile-based renderer, allowing competitive training times compared to SOTA fast radiance field methods. Once trained, our renderer allows real-time navigation for a wide variety of scenes.
	}
	\label{fig:overview}
\end{figure*}

\section{Triplet}
\label{sec3}
Triplet is designed to enable an simulation of scene's parameter including geometry, materials and lighting. Different from scene optimization where meshes which usually are required to be watertight (e.g. Sphere Mesh) or provision of accurate geometry, it would be much favorable to lower the threshold of initialization for scene or object with complex, non-convex geometry characteristic, where many simulation techniques for triangle mesh could be easily plugged in. For these purpose, we start our optimization with sparse point and assemble them into non-watertight and scattered triangle faces binding with materials and lighting. 

Triplet is based on triangle mesh but allow for more flexibility during optimization. Therefore \textbf{triplet} can be represented by four main sets: 

\begin{itemize}
    \item \textbf{Vertices} (\(V\)): A collection of points in 3D space, where each vertex is a tuple representing its coordinates:
    \[
    V = \{v_1, v_2, \dots, v_n\}, \quad v_i = (x_i, y_i, z_i)
    \]
    \item \textbf{Vertex Property} (\(P\)): A collection of RGB values binding at vertex, where each vertex is a tuple representing its coordinates:
    \[
    V_p = \{\mathbf{p_1},\mathbf{p_2}, \dots, \mathbf{p_n}\}, \quad p_i = (material_i,texture_i, alpha_i)
    \]
    \item \textbf{Edges} (\(E\)): A collection of unordered pairs of vertices, each representing a connection (edge) between two vertices:
    \[
    E = \{(v_i, v_j) \mid v_i, v_j \in V \text{ and } v_i \text{ is connected to } v_j\}
    \]
    Each edge \( (v_i, v_j) \) is an unordered pair of vertices.
    
    \item \textbf{Faces} (\(F\)): A collection of triangles, where each triangle is defined by an ordered triplet of vertices:
    \[
    F = \{(v_i, v_j, v_k) \mid v_i, v_j, v_k \in V \text{ and form a triangle}\}
    \]

\end{itemize}
To allow optimization for complex geometry, triplet does not requires mesh to be closed and manifold during early optimization. Also, triplet conduct alpha blend to increase the visibility of all the triplet in the scene. For the ith visible triplet with alpha value $\alpha_i$, it reflect/refract $alpha_i$ of the energy $E_i$, and the energy will be absorb by the $(i-1)_{th}$ triplet occluding in screen space . Therefore, for the $i_{th}$ triplet, the energy remaining is:
\begin{equation}
    E_r = 1 * \prod_{j=1}^{i-1} (1 - \alpha_j).
\end{equation} and the energy showing at pixel is:
\begin{equation}
    E =\sum_i \alpha_i * E_i * E_r = E_i * \alpha_i* \prod_{j=1}^{i-1} (1 - \alpha_j).
\end{equation}Usually we use RGB value to visualize the energy E.

\subsection{Optimization}
The optimization of triplet include two different mode which is the optimization of discrete meshes and the connected mesh. 
The most important part of our algorithm is how to create a appropriate cover on the surface, the optimization on triplet help to achieve this object. While we optimize the Geometry to get a smoother and simpler mesh aligned to the potential surface. During the optimization of discrete triplet, we adopt optimization strategy inspired by gaussian splatting\cite{gs3d}, including gradient-based optimization and adaptive density control but different in the details of implementation and the tuning of surface with closed and manifold meshes. 

For materials such as albedo, metallic, ambient occlussion and alpha, we use Sigmoid function to constrain the parameter into range [0,1], which provides a fluent gradient during optimzation
Triplet optimize scene with multi-view RGB input. We use image loss to optimize parameters:
\begin{equation}
    L_{c} = L_1 + L_{ssim}
\end{equation}
We also regulate the process with total variation on image:
\begin{equation}
    L_{ITV} = \sum_{i,j} \sqrt{(x_{i,j-1} - x_{i,j})^2 + (x_{i+1,j} - x_{i,j})^2}
\end{equation}
As a physically-plausible simulation, our method naturally give good estimation of normal accompanying with light, where the normal and light affect each other in the progress of optimization. To accelerate this spiral optimization process, we further regularize optimization with normal consistency loss, where for discrete triplet:
\begin{equation}
    L_n = \sum_i \omega_i (1 - n_{i}^T \cdot N),
\end{equation}

For optimization of connected triplet which becomes meshes with good manifold property, we use loop divison to densify the mesh and also simplify the mesh with Quadric Error Metrics. When extracted mesh from triplet, we use assign vertex property with nearest-neighbor. Also, we filter vertex materials in ring neighborhood during optimization by average the materials and calculate the total variation in 1-ring neighbor, where:
\begin{equation}
    L_{GTV} =  \sum_i \sum_{j \in \mathcal{N}_i^1} \sqrt{W_{ij}} \|x_i - x_j\|_1,
\end{equation}

Also we applied normal consistency regulation on connected mesh. For neighboring faces $f_0, f_1$ in connected triplet:
\begin{equation}
    L_{n}(f_0, f_1) = 1 - cos(n_0, n_1),
\end{equation} where $cos(n_0, n_1) =\frac{n_0\cdot n_1 }{ ||n0||*||n1||}$ is the cosine of the angle between the normals $n_0$ and $n_1$, and $n_0 = (v_1 - v_0) \times (a - v_0), n_1 = - (v_1 - v_0) \times (b - v_0) = (b - v_0) \times (v_1 - v_0).$ 
We smooth the mesh with laplacian loss\cite{laplaciansmooth}:
\begin{equation}
     L_{V_i} = \sum_j w_{ij} (v_j - v_i).
\end{equation} We use uniform variant,where $w_{ij} = 1 / |S_i|$ and $S_i$ is laplacian matrix. 

\textbf{Density control for discrete and connected meshes} The most fundamental steps for further optimization of other factors is how to create dense cover on surface with triplet which allow gradient to propagate. To achieve this target,  we adopt similar strategy as Gaussian Splatting who control the density of primitives with gradient and size as indicator. Different from gaussian splatting, the gradient of triplet consist of three gradient of vertex, where the gradient accumulation could be quite different between different vertex, especially at edge cases. There are many reasonable method to divide the faces into different area which offer a good division of area where a clear division of gradient could be achieve. We use loop division for splitting which divides face into four part. Loop division offer a good quarantine between high-gradient area and low-gradient area. When clone triangle face, we clone the face and move the vertex toward gradient direction. 

\textbf{Density control for connected meshes} For connected meshes, we also use gradient as an information for simplification or densification. For area with large gradient accumulation, we increase the density of mesh . For area with small gradient accumulation, we simplify the mesh.

\section{Experiment}
\subsection{Implementation}
We provide a PyTorch-based implementation for evaluating our method. However, PyTorch may be less efficient in terms of speed and VRAM consumption, particularly when dealing with memory-intensive operations such as rasterization, ray tracing, and the computation of complex materials. To address this, we also offer CUDA-based implementations for certain operations. Specifically, we leverage the differentiable rasterizer from PyTorch3D\cite{ravi2020pytorch3d}, and for more customizable shaders, we use PyCUDA, which offers both the flexibility for shader programming and efficiency comparable to native CUDA. For operations requiring modification, such as Spherical Harmonics (SH) lighting, we provide a fully CUDA-based implementation.

\textbf{Optimization Details}
Rasterization: For efficiency, we rasterize up to 30 faces per pixel during the main optimization. To clear noisy triplet initializations, we warm up the optimization process at a lower resolution, rendering up to 150 faces per pixel initially. The optimization begins with images at 1/4th the target resolution and 150 faces per pixel, then progressively upsamples the resolution. Specifically, we double the resolution after 200 and 600 iterations. After the first 200 iterations, we reduce the number of faces per pixel to 30 for the remainder of the training.

For vertex-based SH lighting, we initialize the SH coefficients with random directions and a radiance of 1. During rasterization, we begin training with the zero-order SH component and gradually increase the SH band by 1 every 1,000 iterations. In contrast, for ray tracing, we optimize the full SH band from the start. We adopt a total SH band of 5, commonly used in image-based lighting (IBL). To ensure stable training, we clip the gradient norm of the SH coefficients to a maximum value of 1. For image based lighting, we use SH band of 9. For point light, we initialize the location at the center of the scene with intensity 40 and white light. For directional light, we initialize with direction (0,1,0) and intensity of 10.

\subsection{Qualitative Results}
We test our method with Mip-NeRF360 dataset, Blender Synthetics dataset. All the experiments share the same initialization setting. We conduct all experiment with single RTX-4090 GPU.  Results are shown in Fig.\ref{fig:overview} and Fig.\ref{fig:teaser}

\begin{figure}[!h]
	\includegraphics[width=0.49\columnwidth]{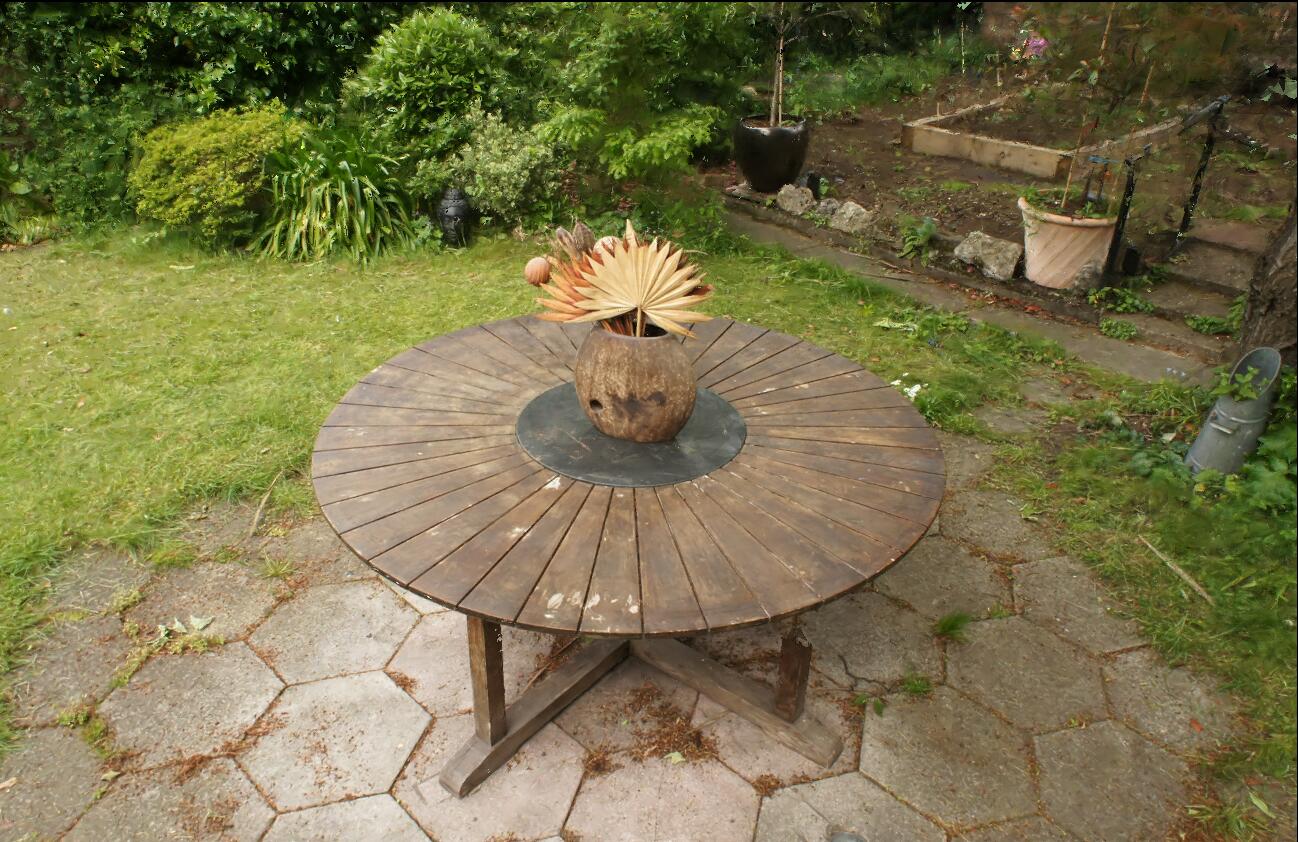}
	\includegraphics[width=0.49\columnwidth]{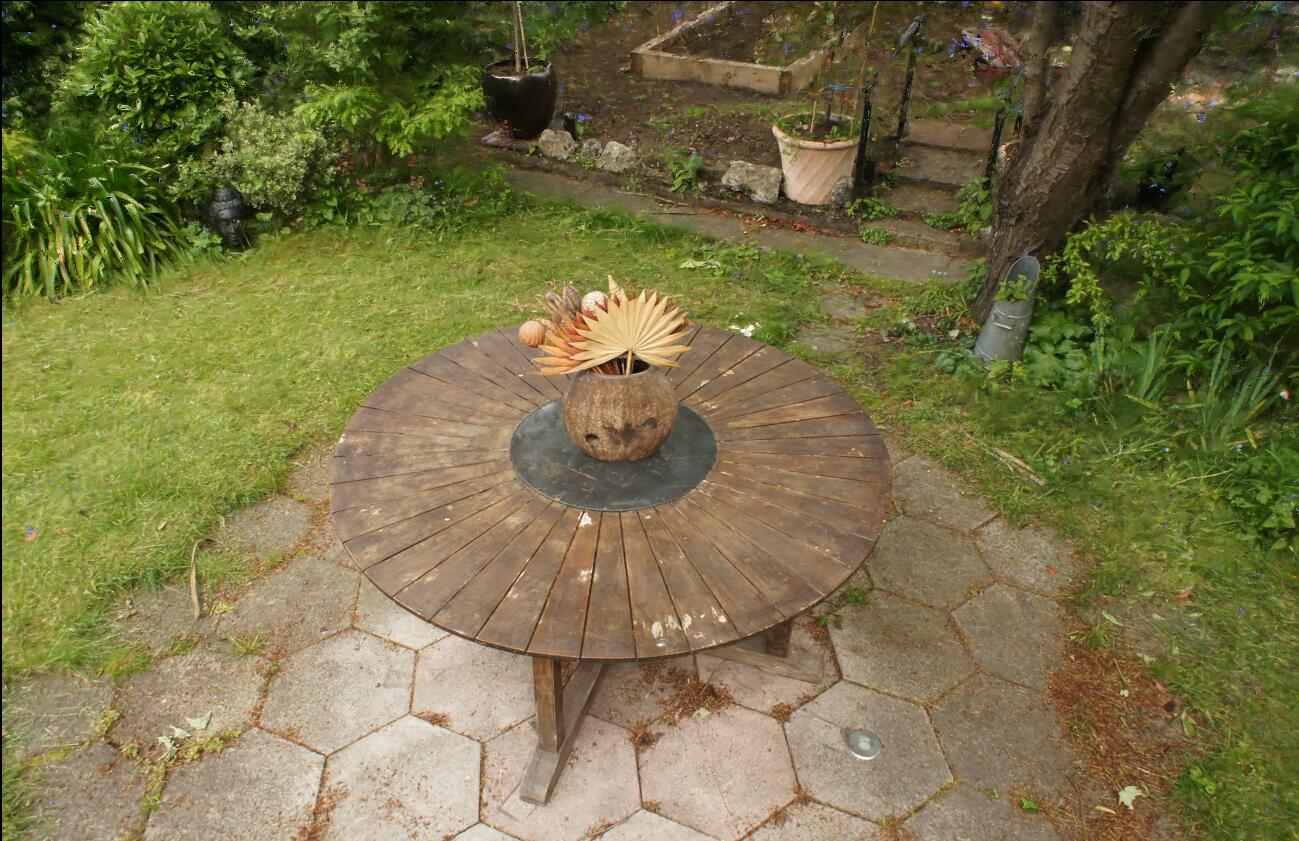}
	\includegraphics[width=0.49\columnwidth]{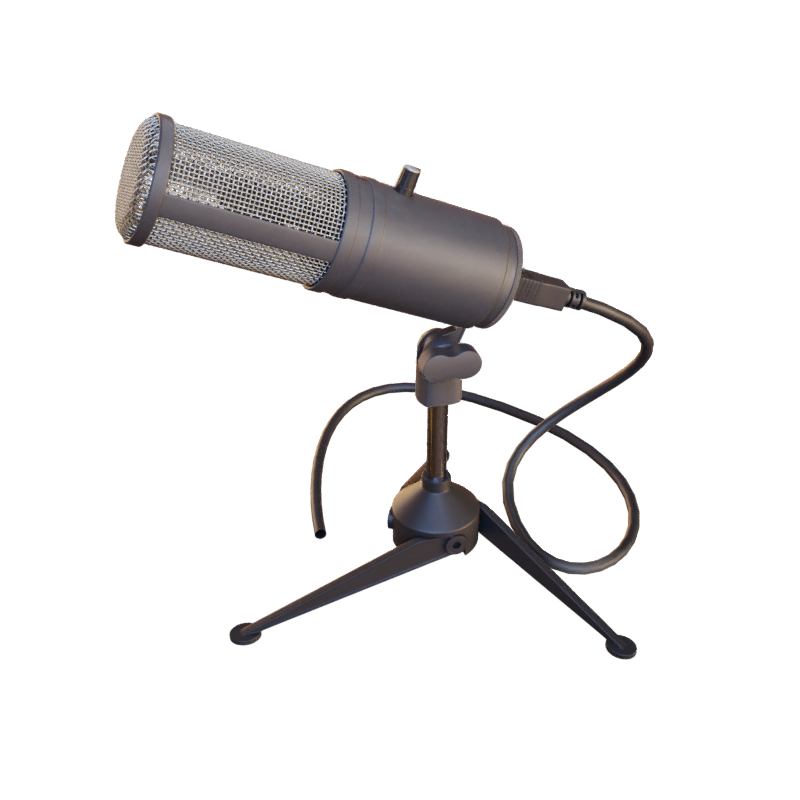}
	\includegraphics[width=0.49\columnwidth]{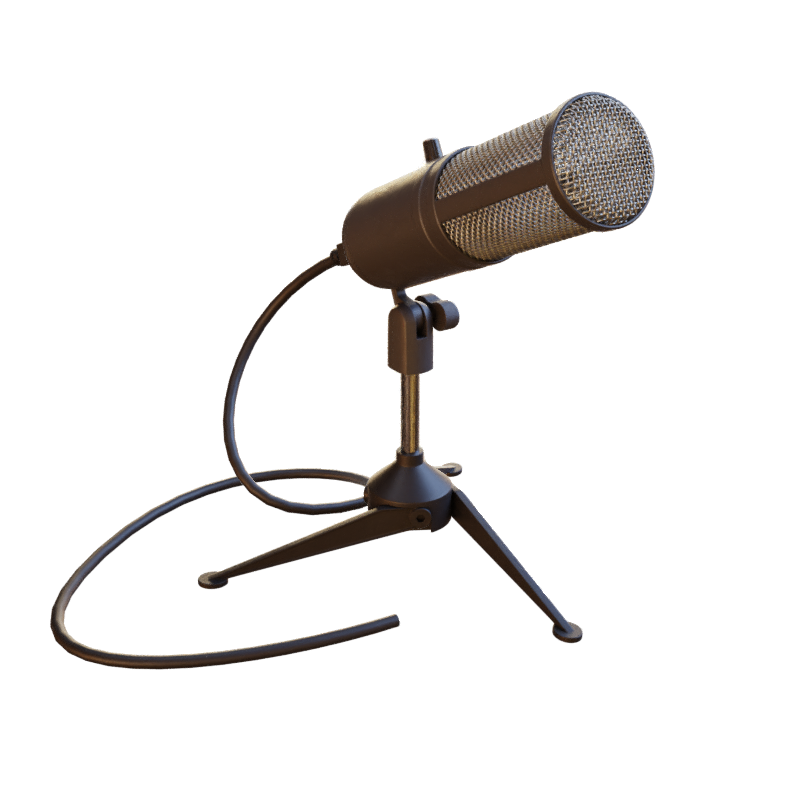}
	\caption{
		\label{fig:limit}
 Triplet can simulate realistic light sources, materials, and their interactions, introducing view-dependent effects such as multiple light sources from the environment and inter-object reflections.\textsc{Train} scene.
	}
\end{figure}

\section{Conclusion and Future Works}
\textbf{Conclusion}
In this paper, we introduced Triplet, a novel framework designed to enable photorealistic novel-view synthesis and inverse rendering using multi-view RGB images as input. The key challenges addressed in this work were creating a dense cover over potential surfaces and effectively utilizing triangle meshes to fit complex geometries. We adopted a gradient-based optimization strategy with adaptive density control, allowing each triangle face to act as a flexible unit. This flexibility makes Triplet more suitable for handling complex geometry compared to traditional mesh-based methods.

As a mesh-based approach, Triplet leverages the well-established rendering pipeline from computer graphics, particularly physically-based rendering (PBR), to create realistic applications. This adaptability with CG techniques significantly increases the potential impact of Triplet in various real-world applications.

\textbf{Limitations and Future Work}
While Triplet demonstrates significant potential, there are limitations to be addressed. The interaction between materials and lighting is critical for accurate simulation, and Triplet struggles to capture material properties that are not directly observable. To mitigate this, we applied filtering and propagation techniques, assuming local stability in material properties, but this is a partial solution. It would also be interesting to see how advanced graphics technique could be adapted into Triplet, such as photon mapping.

A promising future direction would be to explore how neural networks, particularly Graph Neural Networks (GNNs), could assist in resolving this challenge on inferring the materials of unobservable part in the scene. Additionally, Triplet does not currently handle sophisticated materials, especially anisotropic materials (those with directionally dependent reflections), which are essential for simulating effects such as hair or glossy surfaces, although it could be approximate by more faces. Extending Triplet to accommodate these materials would be a valuable enhancement in both accuracy and efficiency.

Lastly, extracting lighting priors directly from input images could further improve the realism of the simulations. Investigating techniques to infer lighting information from input data would be an exciting avenue for future research. An AutoEncoder accompanying with Image-based lighting might work well.

\bibliographystyle{ACM-Reference-Format}
\bibliography{sample-base}

\end{document}